\documentstyle[12pt]{article}
\renewcommand{\aa}{{\cal A}}

\begin{document}
\font\ninerm = cmr9

\def\footnoterule{\kern-3pt \hrule width \hsize \kern2.5pt}

\pagestyle{empty}
\begin{center}
{\large\bf On the area operators of 
the Husain-Kucha\v r-Rovelli model
and Canonical/Loop Quantum Gravity}
\end{center}
\vskip 1.5 cm
\begin{center}
{\bf Giovanni AMELINO-CAMELIA}\\
\end{center}
\begin{center}
{\it Institut de Physique, Universit\'e de Neuch\^atel,
rue Breguet 1, Neuch\^atel, Switzerland}\\

\end{center}

\vspace{1cm}
\begin{center}
{\bf ABSTRACT}
\end{center}

{\leftskip=0.6in \rightskip=0.6in

I investigate the relation between an 
operative definition of the area of a surface
specified by matter fields
and the 
area operators recently introduced in
the canonical/loop approach to Quantum Gravity
and in Rovelli's
variant of the Husain-Kucha\v r Quantum-Gravity toy model.
The results 
suggest
that the discreteness of the
spectra of the area operators 
might not be observable.

}

\vskip 3.4cm
%
%

\vfill

\noindent{
NEIP-98-003 \space\space\space 
gr-qc/9804063
\hfill April 1998}

\newpage
\baselineskip 12pt plus .5pt minus .5pt
\pagenumbering{arabic}
\pagestyle{plain} 


One of the most intriguing aspects
of Canonical/Loop Quantum Gravity~\cite{canoni,loop,loopash}
is its area operator~\cite{arsarea} which has discrete eigenvalues.
Although in the context considered in Ref.~\cite{arsarea} areas
are not diffeomorphism-invariant,
the analysis reported by Rovelli in Ref.~\cite{rovarea}
suggests that a discrete spectrum should also
characterize areas specified
in a diffeomorphism-invariant
manner~\cite{mrsold,mrsrov} by matter fields.
In fact, in Ref.~\cite{rovarea} 
this discreteness was analyzed within
the model obtained by introducing matter fields
in the Husain-Kucha\v r quantum-gravity toy model~\cite{huskuch},
whose area operator is completely analogous to the one of
Canonical/Loop Quantum Gravity.
In this brief note, I shall assume that indeed such a discrete
area operator correctly describes areas in the quantum-gravity
formalism, and investigate how its mathematical properties
would affect the outcome of experiments in which areas are
measured. 
This issue was only very briefly considered in Ref.~\cite{rovarea}.

Some of the points made in the following are relevant for the
study of any diffeomorphism-invariant area operator
(whether or not the spectrum is discrete). Other aspects of 
the analysis apply only to diffeomorphism-invariant area operators
with discrete spectrum, but still the details of the spectrum
are never important for the line of argument here proposed.
For simplicity, the reader can assume that the area operator 
has eigenvalues $\aa_n$ given by half-integer multiples
of the square of the ``Planck length'' $L_P$ ($L_P \sim 10^{-33}cm$):
\begin{eqnarray}
\aa_n = {n \over 2} L_P^2 ~,
\label{eigen}
\end{eqnarray}
which is the type of quantization found~\cite{rovarea}
in the Husain-Kucha\v r-Rovelli model.

Of course, it will be here necessary to analyze
a procedure for the measurement of areas.
I shall consider the procedure proposed by Rovelli in 
Ref.~\cite{rovarea}. There, for simplicity, the matter fields 
that specify the surface whose area is being measured
are taken to form a metal plate, and the area $\aa$ of this metal plate is 
measured using an electromagnetic device that keeps a second metal 
plate at a 
small distance $d$ and measures the capacity $C$ 
of the so formed capacitor.
Of course, measuring $d$ and $C$, and assuming 
that $d \ll \sqrt{\aa}$, one also measures $\aa$ as 
\begin{eqnarray}
\aa = C d ~,
\label{adc}
\end{eqnarray}
where I chose for simplicity units in which the 
relevant permittivity is 1.

In a conventional Quantum Mechanics context 
one can establish with total accuracy the properties
of the spectrum of a given operator in the limit in which
the devices composing the measuring apparatus ``behave classically.''
In fact, at the price of renouncing any information
on a conjugate operator, in this classical-device limit 
(Copenhagen interpretation) one can 
in principle measure any given observable
with total accuracy (see, {\it e.g.}, Ref.~\cite{wign}). 
In such a limit one would for example uncover the nature of
the discrete spectra of the area operators of interest here.
In this note I shall investigate
the implications for the measurability of the properties 
of discrete area operators of the fact that, 
as already observed 
in Ref.~\cite{gacmpla}, the classical-device limit is not 
consistent with the nature of the 
gravitational interactions.

In order to illustrate in which sense the classical-device limit is 
not available in Quantum Gravity it is useful to briefly review
the analysis reported in Ref.~\cite{gacmpla}, which focused on the
measurability of the distance $L$ between (the centers of mass of)
two bodies.\footnote{This 
length observable is of course diffeomorphism-invariant
since the two bodies physically identify the points whose distance
is being measured.}
In Ref.~\cite{gacmpla} 
the distance $L$ is measured
via the Wigner measurement procedure~\cite{wign,ng},
which relies on the exchange of a signal between the two bodies.
The setup of the measuring apparatus schematically 
requires {\it attaching}\footnote{\tenrm
\baselineskip = 14pt
Of course,
for consistency with causality,
in such contexts one assumes devices to be ``attached non-rigidly,''
and, in particular, the relative position
and velocity of their centers of mass satisfy the standard 
uncertainty relations of Quantum Mechanics.} 
a light-gun ({\it i.e.} a device 
capable of sending
a signal when triggered), a clock, 
and a detector to one of the bodies
and {\it attaching} a mirror to the other body.
By measuring the time $T$ needed by the signal
for a two-way journey between the bodies one 
also
obtains a 
measurement of  $L$.
[For example, in Minkowski space 
and neglecting quantum effects 
one simply finds that 
$L = c {T / 2}$, with $c$ denoting the speed of light.]
Within this setup it is easy to realize that 
$\delta L$ can vanish only if 
all devices used in the measurement behave classically.
One can consider for example the
contribution to $\delta L$ coming from 
the uncertainties that affect the relative motion of the clock
with respect to the center of mass of the system 
composed by the light-gun and the detector.
This relative position is crucial for the measurement 
procedure since it is associated to two time delays
that must be taken into account in extracting a measurement of $L$.
The first time delay occurs initially when the clock triggers the light-gun
and the second time delay occurs in the end when (having collected 
the ``return signal'') the detector stops the clock.
Let us denote with $x^*$ the relative position of the clock
with respect to the center of mass of the system
composed by light-gun and detector, and use $v^*$ to denote
the corresponding relative velocity.
It is easy to show \cite{wign,gacmpla,ng} that
the uncertainties $\delta x^*$ and $\delta v^*$
that characterize the state in which the 
experimentalist prepares the devices
contribute to $\delta L$ according to
\begin{eqnarray}
\delta L \geq 
\delta x^* + T \delta v^* 
\geq 
\delta x^* 
+ { (M_c + M_{l + d}) \over 
2 M_c \, M_{l + d} } { \hbar T \over \delta x^* } 
~,
\label{dawign}
\end{eqnarray}
where $M_c$ is the mass of 
the clock, $M_{l+d}$ is the total mass of the system composed of
the light-gun and the detector, 
and the right-hand-side relation
follows from observing that Heisenberg's {\it Uncertainty Principle} 
implies $\delta x^* \delta v^* \ge \hbar (M_c + M_d)/(2 M_c M_d)$.
Clearly, Eq.~(\ref{dawign}) implies that $\delta L = 0$ can
only be achieved in the ``classical-device limit,'' understood
as the limit of infinitely large $M_c$ and $M_{l+d}$.
This is consistent with the nature of the ordinary Quantum-Mechanics
framework, which relies on classical devices.
However, once gravitational interactions are taken into account
the classical-device limit is no longer available.
Large values of the masses $M_c$ and $M_{l+d}$ necessarily lead
to great distorsions of the geometry, and well before 
the $M_c , M_{l+d} \! \rightarrow \! \infty$
limit the Wigner measurement procedure can no longer be
completed. 
[For large enough masses we even expect that ``information
walls'' (the ones of black-hole physics) would form between
the elements of the measurement procedure.]

Since the classical 
limit $M_c , M_{l+d} \! \rightarrow \! \infty$ is not 
available,
from Eq.(\ref{dawign})
one concludes that 
in Quantum Gravity the uncertainty 
on the measurement of a length grows with 
the time $T$ required by the measurement procedure
(as it happens in 
presence of decoherence effects \cite{karo}).
In fact, 
from Eq.(\ref{dawign}) one arrives \cite{gacmpla}
at a minimum uncertainty
for the measurement of a distance $L$ of the 
type\footnote{Besides the uncertainties introduced by the devices 
there should also be a measurement-procedure-independent
contribution $L_{QG}$ to this uncertainty.
In most Quantum-Gravity scenarios $L_{QG}$ is identified with
the Planck length~\cite{padma}, whereas in String Theory $L_{QG}$ is the
string length~\cite{venezkonish}. 
I am here for simplicity not keeping track of
this ``minimum length,'' whose implications (possibly involving
non-locality~\cite{alu})
are by now well
accepted. 
From the results of Refs.~\cite{gacmpla,qgess98},
where the measurability of distances was discussed taking
into account both the uncertainty introduced by the ``non-classical'' 
devices and the uncertainty associated to the minimum length,
the reader can easily realize that $L_{QG}$
would not affect the line of argument
here presented.}
\begin{eqnarray}
minimum \left[ \delta L \right] \, \sim \, 
\sqrt{{ c T L_{QG}^*}} \,
\sim \, 
\sqrt{L \, L_{QG}^*}
~,
\label{gacup}
\end{eqnarray}
where $L_{QG}^*$ is a 
Quantum-Gravity length scale
that characterizes the above-mentioned limitations
due to the absence of classical devices,
and the relation on the right-hand side follows from
the fact that $T$ is naturally proportional~\cite{gacmpla,ng} 
to $L$.
Although $L_{QG}^*$ emerges in a way that does not appear
to be directly related to
the Planck length, it seems plausible \cite{gacmpla} 
that $L_{QG}^* \sim L_{P}$.

Having clarified in which sense classical devices are not
available in Quantum Gravity, and having briefly reviewed how this affects 
the measurability of distances, we can now return to the analysis
of  the limitations
on the measurability of the properties 
of discrete area operators.
According to Eq.~(\ref{adc}), in general 
the uncertainty in the measurement of the area $\aa$
receives contributions from uncertainties in the determination
of $C$ and $d$.
Since I am aiming for a final result formulated as a measurability bound
({\it i.e.} a lower bound on the uncertainty),
it is legitimate to ignore the contribution coming from the
uncertainty in $C$ and focus 
on the contribution coming from the uncertainty in $d$
\begin{eqnarray}
\delta \aa \, \ge \, C \, \delta d \, = \, {\delta d \over d} \, \aa ~,
\label{deltadc}
\end{eqnarray}
where I also used again Eq.~(\ref{adc}) to eliminate $C$.

Based on the bound (\ref{gacup}) on the measurability of distances 
one can assume that $\delta d /d \ge \sqrt{L_{QG}^*/d}$ and therefore
\begin{eqnarray}
\delta \aa \, \ge \, \sqrt{L_{QG}^*} \, {\aa \over \sqrt{d}}~.
\label{deltadc2}
\end{eqnarray}
This relation confronts us with a scenario similar to the one of
Eq.~(\ref{dawign}). 
It formally admits a 
limit ($d \rightarrow \infty$) in which
the area could be measured with complete accuracy, but this limit
cannot be reached within the constraints set by
the nature of the measurement procedure.
In fact, the relation (\ref{adc}), on which the measurement 
procedure is based,
only holds for $d \ll \sqrt{\aa}$, and in considering larger and larger $d$
one quickly ends up loosing all information on $\aa$.
A rather safe lower bound is therefore 
obtained by imposing $d \le \sqrt{\aa}$
in Eq.~(\ref{deltadc2}), which gives
\begin{eqnarray}
\delta \aa \, \ge \, \sqrt{L_{QG}^*} \,\, \aa^{3/4} ~.
\label{deltadc3}
\end{eqnarray}
Interestingly, Eq.~(\ref{deltadc3}) is, like Eq.~(\ref{gacup}),
the result of the uncertainties in the position of 
a device, in this case the metal plate used for the measurement.
However, Eq.~(\ref{deltadc3}) 
was not derived by observing that the limit 
of infinitely-massive metal plate
is not available.
Rather than resulting from the properties of the metal plate,
the bound (\ref{deltadc3}) follows from the general limitations
on the measurability of distances encoded in Eq.~(\ref{gacup}).
Of course, a more detailed analysis of this measurement procedure
would have to take into account also the properties of 
the metal plate. While I shall not here attempt such a delicate analysis,
it is perhaps worth emphasizing some of its elements of difficulty.
The fact that ideal measuring plates (just like ideal measuring 
rods~\cite{wign,gacmpla,ng}) would not be available in Quantum Gravity 
can be discussed very simply by viewing
a plate as composed by elementary cells,
possibly of size $L_P^2/2$ as encoded in
area-quantization relations of the type (\ref{eigen}).
In ordinary (non-quantum) gravitational contexts one would obtain
such a plate by setting in rigid motion
the elementary cells that compose it;
however, once quantum effects are switched on 
the {\it Uncertainty Principle} does not allow
to maintain fixed during the measurement the relative position of the 
elementary cells composing the plate, thereby
excluding the possibility of rigid motion. 
This Quantum-Gravity description\footnote{It is perhaps
worth emphasizing that ideal measuring plates
(like other ideal classical devices)
are consistent with the laws of ordinary
(non-gravitational) Quantum Mechanics.
In fact,
in the limit in which
each elementary cell has infinite mass the {\it Uncertainty Principle}
ceases to affect the dynamics of the cells and therefore the
cells can (at least in principle) 
be set in rigid motion with respect to one another.
This infinite-mass classical limit is perfectly consistent
with the conceptual structure of Quantum Mechanics and with the
non-gravitational analysis of measurement procedures,
but, as explained above, it is not consitent with the nature
of measurement procedures once gravitational interactions are turned on.}
of a plate
of course introduces 
new elements of uncertainty in the analysis of the measurability
of areas measured using the procedure considered in
the present note. Having ignored this
(which is probably the
most significant) source of limitations for the measurability,
we can expect that the actual measurability bound
for areas in Quantum Gravity could be much tighter than
the bound (\ref{deltadc3}).

While even tighter measurability 
bounds might be uncovered by more refined
analyses, already the 
bound (\ref{deltadc3}) 
appears to require a significant
shift in the
physical interpretation of 
area-quantization relations of the type (\ref{eigen}).
In Ref.~\cite{rovarea} it was observed that the formal
property (\ref{eigen}) of the area operator in the 
Hilbert space of the Husain-Kucha\v r-Rovelli model
(and similar considerations should apply to the area operator
of Canonical/Loop Quantum Gravity)
would directly affect the outcome 
of area mesurements 
within the conventional
Quantum Mechanics framework, {\it i.e.} 
the outcome of area measurements should be $L_P^2/2$ quantized.
I have here observed that the conventional 
Quantum Mechanics framework, with its classical 
measuring apparatus, is inconsistent with the nature of
gravitational interactions and that the
heuristic analysis of a new Quantum Gravity measurement
framework appears to suggest that the area quantization encoded
in the formalism {\it might not be observable}.
In fact, assuming $L_{QG}^* \sim L_P$, Eq.~(\ref{deltadc3}) 
indicates that the measurement of a given area of order $n  L_P^2/2$
would be affected by an uncertainty of at least $\sim L_P^2 (n/2)^{3/4}$,
{\it i.e.} (for every area with $n > 1$)
an uncertainty much larger than the $L_P^2/2$ quanta.

Concerning the physical interpretation of Eq.~(\ref{deltadc3}) 
one is also naturally led to inquire about the type of symmetries 
that could result in such a structure. Of course, 
it will be possible to rigorously address this question only
once a formalism supporting relations such as
(\ref{gacup}) and (\ref{deltadc3}) is found;
however, some consistency arguments~\cite{qgess98,kpoinpap}
appear to indicate that dimensionful deformations of 
Poincar\'e symmetries might be involved.
While I shall not repeat those arguments here, it is worth emphasizing,
as a reason of interest in the scenario advocated in the present note,
that such deformations of Poincar\'e symmetries
could soon be tested \cite{grbgac}
experimentally by exploiting the recent dramatic 
developments in the phenomenology of gamma-ray bursts \cite{grbnews}.

In closing, 
let me summarize the points made in this note 
also clarifying which ones could be considered 
as ``robust.''
The way in which
the new bound (\ref{deltadc3}) has been here derived
involves rather heuristic arguments, and 
might reflect the structure of
the specific example of procedure for the measurement of areas
which has been considered. Accordingly, 
Eq.~(\ref{deltadc3})
is to be considered as very preliminary, and in particular
more refined analyses might find that the $\aa$-dependence 
on the right-hand-side comes in with an exponent different from $3/4$.
However, Eq.~(\ref{deltadc3}) should be expected
to capture the correct qualitative behavior, {\it i.e.} a limitation
on area  measurability that grows with the size of the area.
In fact, such a behavior already characterizes measurements in 
ordinary Quantum Mechanics, unless
the infinite-mass ``classical-device'' limit is taken.
The observation that this ``classical-device'' limit is not available
once gravitational interactions are taken into account
is nearly self-evident and has been here discussed rather intuitively.
I emphasize that this observation is not in contradiction
with the point made by some authors (see, {\it e.g.}, Ref.~\cite{rovarea})
that even in Quantum Gravity the measuring apparatus
should be ``external''
to the system under observation.
Indeed, I have analyzed the measuring devices as external
to the system, {\it i.e.} I maintained at all stages
the distinction between the degrees of 
freedom of the system being observed and the ones of the measuring
apparatus.
The novel element of the analysis here reported
is that I refrained from
assuming that these external devices would somehow
not be subject to the same laws of physics that
govern the dynamics of the system under observation.
This assumption is not made in ordinary non-gravitational
Quantum Mechanics\footnote{As illustrated by some of the 
points made in 
this note ({\it e.g.} the different descriptions that Quantum Gravity
and ordinary Quantum Mechanics give of a measuring plate), 
in ordinary non-gravitational
Quantum Mechanics by considering measuring devices 
that behave classically one is not actually assuming
that the laws of Quantum Mechanics would not apply to these
devices; in fact, the classical limit (typically involving
an infinite-mass limit) is perfectly well defined
within ordinary Quantum Mechanics and leads to no pathologies
as long as the gravitational interactions are ignored.}
and there appears to be no reason why it should be made
in the Quantum-Gravity context.
Assuming this point is correct
Eq.~(\ref{deltadc3}) should 
capture the correct qualitative structure of 
the area measurability bound in Quantum Gravity,
and consequently,
as explained above,
one expects that the output of area measurements
could not take the form of integer multiples
of the $L_P^2/2$ quanta even though such a quantization
charaterizes the formal spectrum of the area operator.

Besides considering other procedures for the measurement
of areas and refining the analysis of the implications 
of the non-classical behavior of devices,
future work aiming at establishing more precisely the form
of the area measurability bound in Quantum Gravity
should also consider how the {\it Equivalence Principle}
could affect the measurement procedures.
This might play a crucial role in the way the devices
interact with the system being observed.
The conceptual framework of ordinary Quantum Mechanics
relies not only on the infinite-mass ``classical-device'' limit,
but also on the limit in which the devices decouple
from the system.
For example the devices used
in the measurement of the electromagnetic field
interact with it, but (as 
emphasized in Refs.~\cite{qgess98,rose,bergstac})
in measurability analyses within ordinary Quantum Mechanics
it is crucial that there is a
limit in which devices fully decouple from the field.
In the case of the electromagnetic field this limit
is the one in which the devices have vanishing
ratio of electric charge to inertial mass. 
Of course,
the devices of a Quantum-Gravity apparatus
also interact with the gravitational field,
and the limit in which the devices decouple from the field 
appears not to be available
as a result of the fact
that the ratio of gravitational charge
to inertial mass is fixed by the {\it Equivalence Principle}.
This has not played a role in the present analysis, but could be
important in more refined studies of measurability of area observables
or other Quantum Gravity observables.

%
 

\baselineskip 12pt plus .5pt minus .5pt

\end{document}